\documentclass[floatfix,twocolumn,amsmath,amssymb,aps,prb,showpacs,letter,superscriptaddress]{revtex4}
\usepackage{latexsym,epsfig,bm,times,psfrag,subfigure}

\usepackage{color}

\newcommand{\hto}{Ho$_{2}$Ti$_{2}$O$_{7}$}

\newcommand{\dto}{Dy$_{2}$Ti$_{2}$O$_{7}$}

\newcommand{\dtox}{Dy$_{2-x}$Y$_{x}$Ti$_{2}$O$_{7}$}
\newcommand{\htox}{Ho$_{2-x}$Y$_{x}$Ti$_{2}$O$_{7}$}

\begin{document}


\title{
Non-monotonic residual entropy in diluted spin ice: a comparison between Monte Carlo simulations of
diluted dipolar spin ice models and experimental results}

\author{T. Lin}
\affiliation{Department of Physics and Astronomy, University of Waterloo, Waterloo, ON, N2L 3G1, Canada}
\author{X.  Ke}
\affiliation{Department of Physics and Materials Research Institute,
Pennsylvania State University, University Park, PA 16802, USA}
\affiliation{Quantum Condensed Matter Division, Oak Ridge National Laboratory, Oak Ridge,  TN 37831, USA}
\affiliation{Department of Physics and Astronomy, Michigan State University, East Lansing, MI 48824, USA}
\author{M. Thesberg}
\affiliation{Department of Physics and Astronomy, McMaster University, Hamilton, Ontario L8S 4M1, Canada}
\author{P. Schiffer}
\affiliation{Department of Physics and Materials Research Institute, 
Pennsylvania State University, University Park, PA 16802, USA}
\affiliation{Department of Physics, 
The University of Illinois at Urbana-Champaign,
1110 West Green Street, Urbana, IL 61801-3080, USA}
\author{R. G. Melko}
\affiliation{Department of Physics and Astronomy, University of Waterloo, Waterloo, ON, N2L 3G1, Canada}
\affiliation{Perimeter Institute for Theoretical Physics, 31 Caroline North, Waterloo, Ontario, N2L-2Y5, Canada}
\author{M.  J. P. Gingras}
\affiliation{Department of Physics and Astronomy, University of Waterloo, Waterloo, ON, N2L 3G1, Canada}
\affiliation{Perimeter Institute for Theoretical Physics, 31 Caroline North, Waterloo, Ontario, N2L-2Y5, Canada}
\affiliation{Canadian Institute for Advanced Research, 180 Dundas Street West, Suite 1400, Toronto, ON, M5G 1Z8, Canada}

\date{\today}

\begin{abstract}
Spin ice materials, such as \dto $ $ and \hto, have been the subject of much interest for over the past fifteen years. 
Their low temperature strongly correlated state can be mapped onto the proton disordered state of common water ice and, consequently,
spin ices display the same low temperature residual Pauling entropy as water ice.  Interestingly, it was found in a previous 
study [X. Ke {\it et. al.} Phys. Rev. Lett. {\bf 99}, 137203 (2007)] that,  upon dilution of the magnetic rare-earth 
ions (Dy$^{3+}$ and Ho$^{3+}$) by non-magnetic Yttrium (Y$^{3+}$) ions, the residual entropy depends {\it non-monotonically}
 on the concentration of Y$^{3+}$ ions.
 In the present work, we report results from Monte Carlo simulations of site-diluted microscopic dipolar spin ice models (DSIM) 
that account quantitatively for the experimental specific heat measurements, and thus also for the residual entropy, 
as a function of dilution, for both \dtox $ $ and \htox. 
The main features of the dilution physics displayed by the magnetic specific heat data
are quantitatively captured by the diluted DSIM up to, and including, 85\% of the magnetic ions diluted ($x=1.7$). 
The previously reported departures in the residual entropy between 
\dtox $ $ versus \htox $ $, as well as with 
a site-dilution variant of Pauling's approximation,  are thus 
rationalized through the site-diluted DSIM.
For 90\% ($x=1.8$) and 95\% ($x=1.9$) of the magnetic ions diluted, we find a significant discrepancy
between the experimental and Monte Carlo specific heat results.
We discuss some possible reasons for this disagreement.
\end{abstract}


\maketitle

\section{Introduction}

The theoretical and experimental study of geometrically frustrated magnets \cite{Ramirez1994, Diep, Lacroix, Balents2010, Gardner2010} constitutes a very active 
research area in contemporary 
condensed matter physics. In these systems, the predominant interactions compete with each other, inhibiting the development of long-range magnetic order down to very low, if not zero, temperature.~\cite{Balents2010} The temperature regime where strong magnetic correlations exist, but long range order is absent, 
is commonly referred to as spin liquid \cite{Balents2010} or cooperative paramagnetic state.~\cite{Villain1979}

One particularly topical example of geometrically frustrated magnets is spin ice materials.~\cite{Lacroix, Gardner2010, Bramwell_Science2001, Harris1997, Ramirez1999} These are realized by the canonical compounds \dto $ $ and 
\hto $ $, as well as by the less extensively studied Dy$_2$Sn$_2$O$_7$ ~\cite{KeDy2Sn2O7} and Ho$_2$Sn$_2$O$_7$.~\cite{Kadowaki2002} 
More recently, high pressure chemical synthesis has allowed one to make the Dy$_2$Ge$_2$O$_7$ and Ho$_2$Ge$_2$O$_7$ compounds,  and thermodynamic
 measurements have 
shown these materials 
to be an interesting new class of spin ice systems.~\cite{Zhou_Natcomm,Zhou2012} 
In that context, it is interesting to note that the
CdEr$_2$Se$_4$, in which Er$^{3+}$ is unusually described by an Ising spin,
as also been shown to be a spin ice.~\cite{CdEr2Se4}
In all of 
these materials, the magnetic rare-earth Dy$^{3+}$, Ho$^{3+}$ and
Er$^{3+}$ ions sit on the vertices of a pyrochlore lattice of corner-sharing tetrahedra; the Ti$^{4+}$, Sn$^{4+}$ and Ge$^{4+}$ ions are non-magnetic. Because of the very large single-ion anisotropy at play in these systems, the moments can be described below a temperature $T\sim50$ K as classical Ising spins pointing along the local $[111]$ direction 
at their respective pyrochlore lattice sites.~\cite{Rosenkranz2000, Harris1997,Gingras_Springer} 
Below a typical temperature of order 1 K, the magnetic state of (Dy,Ho)$_2$(Ti,Sn,Ge)$_2$O$_7$  can be mapped onto the proton
 disordered state of common water ice,~\cite{Pauling1935} hence the name spin ice.~\cite{Harris1997} 
In this low temperature spin ice state, the magnetic moments are highly correlated locally and obey the so-called ``ice rules'': two spins point in and two spins point out of each tetrahedron of the pyrochlore lattice, but without displaying long range order.~\cite{Bramwell_Science2001} 
The spin ice state can thus be viewed as a cooperative paramagnet,~\cite{Villain1979} 
or a classical spin liquid to adopt a more modern terminology.~\cite{Balents2010} 
The label ``classical spin liquid'' stems from the very strong Ising nature of the 
lowest-energy crystal-field doublet for both Dy$^{3+}$ and Ho$^{3+}$ which results in a dramatically quenched level of 
quantum spin dynamics.~\cite{Gingras_Springer} 
At the same time, the high energy barrier to single spin flips causes the relaxation dynamics to become 
very slow in these materials below  $T\sim 1$ K.
Consequently, spin ices should be viewed as extremely sluggish classical spin liquids.~\cite{Cepas2012}

For water ice, extensive calorimetric studies had been carried out long before 
\cite{Giauque1933,Giauque1936} its magnetic counterparts were discovered.~\cite{Harris1997} 
The nature of the proton disorder in ice was described by Linus Pauling who estimated the 
residual entropy to be $S_{\rm P} = {\text R}/2\ln(3/2)$ per mole of protons 
 (R is the molar gas constant),~\cite{Pauling1935} matching closely 
with  experiments.~\cite{Giauque1933,Giauque1936} 
The same residual entropy is found in spin ice 
materials,~\cite{Ramirez1999,Cornelius2001,Zhou2012,CdEr2Se4,KeDy2Sn2O7,PrandoHo2Sn2O7}
 providing direct thermodynamic evidence for the ice rules being at work. 
In spin ices, the crossover from the paramagnetic phase to the macroscopically degenerate spin ice state with its Pauling residual entropy is signalled by a broad specific heat peak at around $T_{\rm p} \sim 1.2$ K for \dto $ $ \cite{Ramirez1999, Ke2007} and $T_{\rm p} \sim 1.9$ K for 
\hto $ $.~\cite{Bramwell2001} 
There is no thermodynamic phase transition between the high temperature paramagnetic state and the low temperature spin ice state as evidenced by the absence of sharp thermodynamic features at $T_{\rm p}$. 
Theoretical studies have shown that long range magnetostatic dipole-dipole interactions are responsible for the finite entropy
spin ice state in (Ho,Dy)$_2$(Ti,Sn,Ge)$_2$O$_7$ compounds.~\cite{Gingras2001,Isakov2005,Castelnovo2008} 
Yet, it is generally theoretically accepted that the same dipolar
 interactions should give rise to long-range order 
at a critical temperature $T_c \ll T_p$ if true thermal equilibrium could be maintained down to sufficiently low temperature.~\cite{Gingras2001,Isakov2005} 
Indeed, Monte Carlo simulations that employ loop moves to 
generate non-local spin flips, 
which allow the system to remain in thermal equilibrium without violating the ice rules,
 do find a transition to long-range order at low-temperature.~\cite{Melko2001,Melko2004} 
To this date, however, no experiment has found a transition to long-range order in spin ice materials,~\cite{Fukazawa2002}
 presumably because of a dynamical arrest in spin flips \cite{Castelnovo2010} 
and the associated relaxation times growing exponentially fast below a temperature of about 1 K.

Considering the broader context of ice-like systems developing extremely slow dynamics at sufficiently low temperatures, one notes that as water ice is doped with alkali hydroxides, such as KOH or RbOH, a sharp first order transition to long range order occurs at a temperature near 72 K.
At that transition, a large portion of the residual Pauling entropy is released through the latent heat.~\cite{Tajima1982, Tajima1984} 
These experiments suggest that the proton-disordered ice state is somewhat fragile against impurities and that the frustrated disordered ice state with residual entropy can be eliminated through the influence of impurities and/or random disorder.
 Yet, despite much theoretical work, it remains unclear what is the precise mechanism via which 
alkali hydroxides in the water ice system promotes the development of long range order.~\cite{Singer2012}

\begin{figure} [h]
	\begin{center}
	\includegraphics[width=8cm,angle=0]{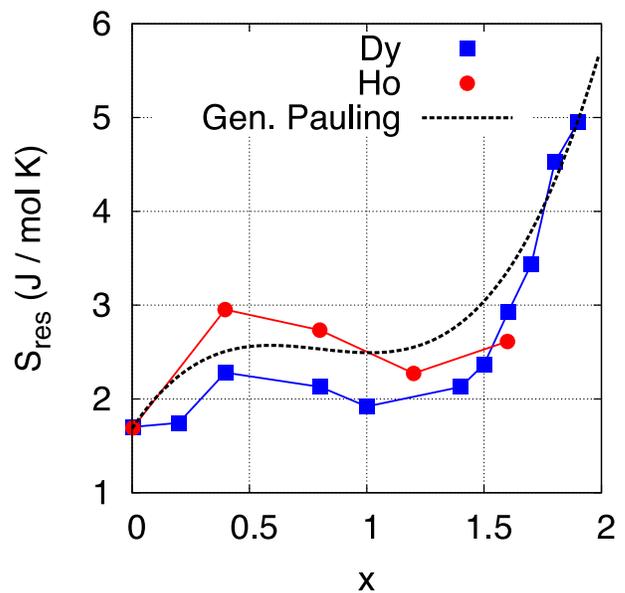}
	\caption{(Color online) Adapted from X. Ke {\it et. al.} Phys. Rev. Lett. {\bf 99} 137203 (2007). 
Experimental residual entropy as a function of dilution level $x$ (see Ref.~[\onlinecite{Ke2007}]). Dy denotes \dtox $ $, 
Ho denotes \htox $ $ and Gen. Pauling denotes the generalized Pauling approximation (gPa) presented in Ref.~[\onlinecite{Ke2007}]. 
As noted by Ke {\it et. al.}, there is an obvious systematic departure between the three curves, 
except for the undiluted compounds ($x=0$).}
	\label{Ke_fig}
	\end{center}
\end{figure}

Inspired by the impurity-driven long-range order observed in water ice,~\cite{Tajima1982,Tajima1984} it is interesting to ask whether the magnetic spin ice analogue could also display interesting behavior when subject to the addition of random impurities.
For example, perhaps a slight dilution of the magnetic Dy$^{3+}$ and Ho$^{3+}$ ions could lower the kinematic barriers for spin flips, thus accelerating the spin dynamics, and help promote a transition to long-range order without significantly affecting the broken discrete symmetry long-range ordered ground state of
dipolar spin ice.~\cite{Melko2001,Melko2004}
In that context, we note that magnetic site-dilution in spin ices can be realized rather straightforwardly
in the \dtox $ $ and \htox $ $ compounds, which form a solid solution over the whole $x \in [0,2]$ range, and where the magnetic Dy$^{3+}$ and Ho$^{3+}$ ions are replaced by non-magnetic Y$^{3+}$ ions.~\cite{Ke2007} 
The close ionic radius of Y$^{3+}$ with that of Dy$^{3+}$ and Ho$^{3+}$ allows for a substitution that causes
negligible local lattice deformation and strain.
Dilution of Dy$^{3+}$/Ho$^{3+}$ by Y$^{3+}$ can thus be viewed as a mere replacement of the 
Dy$^{3+}$/Ho$^{3+}$  magnetic species by a magnetically inert substitute.
 A recent neutron scattering experiment shows no sign of long-range ordering in \htox $ $ 
down to 30 mK for $x=0.3$ and $x=1.0$.~\cite{Chang2010} 
On the other hand, specific heat measurements have found that the low-temperature residual entropy, $S_{\rm res}$, of diluted 
\dtox $ $ and \htox $ $ spin ices display a  non-monotonic dependence on the level of dilution.~\cite{Ke2007} 
A calculation generalizing Pauling's argument \cite{Pauling1935} (gPa) to the case of site dilution  of a nearest-neighbor spin ice model \cite{Harris1997}
was able to qualitatively account for such a non-monotonic behavior.~\cite{Ke2007} 
However, the apparent systematic departures between the gPa and the experiment results 
as well as the differences between Dy- and Ho- based materials (see Fig. \ref{Ke_fig}) have so far remained unaddressed. 
It was suggested in the original work.~\cite{Ke2007}
that the residual entropy may be material-dependent and have a more drastic non-monotonic dependence on levels of dilution than the analytic 
generalized Pauling argument (gPa) does. 
The reason for these differences might be caused, for example, by the extra complexities of the long-range dipolar interactions compared with the nearest-neighbor model.
In this paper, we address and rationalize quantitatively
the origin of the difference in residual low temperature entropy 
between \dtox $ $ and \htox $ $ as well as with the gPa illustrated in Fig.~\ref{Ke_fig}.

Moving away from the specific context of disorder and impurities in ice-like (water or spin) systems, one notes that the problem of quenched random disorder in highly frustrated magnetic systems is one of long-standing interest, going back to the seminal work of Villain.~\cite{Villain1979} 
In more recent years, the effects of disorder on the thermodynamic properties of highly frustrated magnetic
systems, in large part motivated by studies on 
kagome materials such as SrCr$_x$Ga$_{12-x}$O$_{19}$ (SCGO)
 \cite{Ramirez-SCGO,Schiffer-Orphan,Sen-SCGO}
 and  ZnCu$_3$(OH)$_6$Cl$_2$ (Herbertsmithite),~\cite{Mendels-Herbert,Imai-Herbert}
 has been a topic of much interest.
Research efforts in this area have been especially 
 motivated by the necessity to understand whether the observed experimental behavior in SCGO 
and Herbersmithite is intrinsic to the hypothetical disorder-free material or is, instead, (largely) driven by impurity effects. 
The problem of dilution in quantum triangular antiferromagnets is also one with very interesting
 and rich physics.~\cite{Wollny-fractional,Wollny-singular}

Spin ice is at the present time one of the best understood highly frustrated magnetic systems, both from a microscopic model perspective 
\cite{Yavorskii2008}
as well as from a field theory one.~\cite{Henley2005,Henley2010,Sen,Saunders}
Spin ices would thus appear to be an ideal system to investigate 
quantitatively  the effects of random disorder in a highly frustrated magnetic setting.~\cite{Sen,Saunders} 
This is precisely the broader goal of this paper: to perform such a quantitative 
comparison between theoretical modeling and experimental measurements in a specific class of disordered highly frustrated magnetic materials. 
As a first agenda in this program, we consider the aforementioned problem of diamagnetic 
site-dilution dependence of the residual entropy in \dtox $ $ and \htox $ $ spin ice materials. 

It turns out that there is a growing interest in the problem of disorder in magnetic pyrochlore oxides.
For example,  direct \cite{Ross_stuff} and indirect \cite{Revell_NatPhys}
evidence has recently been put forward that, in image furnace grown single crystals, there is a small
level ($O(1\%)$) of substitution of the Ti$^{4+}$ transition metal ions by trivalent rare-earth ions $-$ a phenomenon referred to as ``stuffing''.
Other examples include the mixing of different types of ions on the 
rare-earth site \cite{TbDyTiO} or different non-magnetic ions at the B site.~\cite{YbGaSb,TbSnTiO}
Thus, in comparison with these various disorder settings,
which would all generate random bonds, the problem of site-dilution 
may be expected to be  simpler, and a necessary first step in our goal of understanding the effects of random disorder in magnetic 
pyrochlores oxides.~\cite{Gardner2010}

In order to investigate the microscopic origin of the relative departure of the three curves in Fig.~\ref{Ke_fig}, we performed Monte Carlo simulations of a diluted variant of the pertinent microscopic dipolar spin ice model of \hto \cite{Bramwell2001} and \dto.~\cite{Yavorskii2008} 
A direct comparison of the temperature-dependent magnetic specific heat, $C_{\rm m}(T)$, for various dilution levels, $x$, between simulations and experiments is made in order to validate a simple site-diluted version of the otherwise pure (dilution-free) microscopic models. 
Through the simulation data, we obtain an accurate $C_{\rm m}(T)$, which provides for a precise determination of the residual entropy, down to the lowest temperature $T_0 \sim 0.4$ K considered in experiments.~\cite{Ke2007} The simulation results confirm the previous speculation \cite{Ke2007} that the departure of the 
material-dependent residual entropy from the generalized Pauling argument (gPa) occurs because of material-specific details of the interactions.
That said, our conclusion regarding the difference in residual entropy $S_{\rm res}$ between
\hto $ $ and \dto $ $ is different from the one in Ref.~[\onlinecite{Ke2007}], namely, we find \htox $ $ to have a smaller 
$S_{\rm res}(x)$ than \htox $ $ does.

The rest of the paper is organized as follows: In Section II, we discuss the details of the experimental methods;
In Section III, we present our microscopic models and the Monte Carlo simulation methods; in Section IV, 
we present and discuss the results of the Monte Carlo simulations 
and address the previously reported \cite{Ke2007} material-dependent 
residual entropies along with their departure from the gPa predictions;
Section V concludes the paper.

\section{Experimental Methods and Results}
\label{Sect:exp-method}

Specific heat measurements were performed on Y-diluted spin ice materials, \dtox $ $ and \htox $ $, using a Quantum Design Physical Property Measurement System (PPMS) cryostat with the He3 option via a standard semiadiabatic heat pulse technique. The Dy-based samples were thoroughly mixed with Ag and pressed into pellets to facilitate thermal equilibration.
The scaled Ag specific heat, measured separately, was subtracted from the total specific heat. The phonon contribution was extracted by fitting the data with the Debye formula in the temperature range $T \in [10,20]$ K, and subtracted from the total specific heat to obtain the magnetic specific heat
contribution, $C_{\rm m}(T)$. 
Ho-based samples were pressed directly into pellets and the magnetic specific heat was obtained after subtracting both the phonon and the large Ho nuclear Schottky anomaly contribution.~\cite{Bramwell2001,Cornelius2001} The data, $C_{\rm m}(T)/T$, integrated from $T_0(x)= 0.4 \pm 0.1$ K, depending on the lowest temperature $T_0(x)$ accessed for a given concentration $x$, up to a (`high') temperature $T$,
was used to determine the residual low-temperature entropy, $S_{\rm res}(T_0)$. The previously reported \cite{Ke2007} residual entropy is reproduced here in Fig.~\ref{Ke_fig} for convenience. As discussed in the Introduction, the residual entropy plotted in Fig. 1 varies 
non-monotonically as a function of the Y concentration for both  the \dtox $ $  and the \htox $ $ series, being qualitatively 
captured by a generalization of
Pauling approximation's (gPa) that is represented by the dashed curve.~\cite{Ke2007}

\section{Microscopic Models and Monte Carlo Simulations}

\subsection{Microscopic Models of Spin Ices}

In spin ices, the magnetic moments reside on a pyrochlore lattice, 
which consists of a face-centered cubic lattice of corner-sharing tetrahedra primitive units.~\cite{Gardner2010, Bramwell_Science2001} 
Due to the large energy scale ($\sim 300$ K) of the crystal field splitting between the ground state
doublet and the lowest-energy excited doublet that exist in 
Dy$_2$Ti$_2$O$_7$ and Ho$_2$Ti$_2$O$_7$,~\cite{Gardner2010,Rosenkranz2000,Gingras_Springer} the states that form 
the ground doublet of the Dy$^{3+}$ and Ho$^{3+}$ ions can safely be assumed to be the only thermodynamically 
relevant states below a temperature $T \lesssim 50$ K.

As suggested originally,~\cite{Harris1997} the minimal model that describes the geometrical frustration in spin ices is
\begin{equation}
	\mathcal{H}_{\rm NNSIM} = J_{\rm eff} \sum_{\langle i,j \rangle}  \sigma_i \sigma_j
	\label{Hnn}
\end{equation}
where $J_{\rm eff}>0$ is the effective antiferromagnetic interaction between the $\sigma$'s Ising 
variables. 
This model possesses a Pauling residual entropy, $S_{\rm P}$,~\cite{Oitmaa} 
 and displays at zero temperature an ice-rule obeying ground state characterized by dipolar-like spin-spin correlations that emerge from the ``two-in''/``two-out'' ice rule constraint.~\cite{Henley2005, Henley2010,Sen}

On the other hand, in the real spin ice materials, the Dy$^{3+}$ and Ho$^{3+}$ ions carry a large magnetic moment ($\sim 10$ $\mu_{\rm B}$) and the long range dipolar interactions cannot be ignored.~\cite{Isakov2005, Gingras2001} 
 Given the symmetry of the crystal field ground state,~\cite{Harris1997,Rosenkranz2000} the magnetic moments can be well described by vector spins constrained by the single-ion anisotropy to point strictly parallel or antiparallel to  their respective local [111] direction ({\it i.e.}, along the line from the corners to the centre of each tetrahedron).~\cite{Bramwell_Science2001, Harris1997, Rosenkranz2000} 
Taking the dipolar interaction and the essentially infinite local Ising anisotropy into consideration, the dipolar spin ice model (DSIM)
is defined by the Hamiltonian:
\begin{eqnarray}
\label{gDSM}
\nonumber
{\cal H}_{\rm DSIM} &&=    \sum_{i>j} s_i s_j\,\,
\left\{ \rule{0pt}{18pt}
\; 
\sum_{\nu=1}^{3} 
J_{\nu} \;\; \delta_{r_{ij},r_{\nu}}\; 
{\hat z}_i\cdot{\hat z}_j\; 
\right.
\\
&&
\left.
+\ D { ({r_1}/{r_{ij}}) }^3      \, 
\left[ {\hat z}_i\cdot{\hat z}_j
-3\,({\hat z}_i\cdot\hat{r}_{ij}) ({\hat z}_j\cdot\hat{r}_{ij})
\right]
\rule{0pt}{18pt} \right\}	.
\end{eqnarray}
\noindent where $ \sigma_i = \pm 1 $ are the Ising spin variables. The first term describes the Ising exchange interaction while the second term is the long-range magnetic dipole-dipole interaction. Here,  $\nu = 1, 2$ or $ 3$ refers to first, second or third nearest neighbors respectively,
where $J_{\nu}$ is the exchange coupling and $r_{\nu}$ is the distance between them.
There are two types of third nearest neighbor interactions which we do not differentiate.~\cite{Yavorskii2008}
$\hat{z}_i$ is the local [111] direction of the Ising axis and  $D$ is the strength of the dipolar interactions at nearest-neighbor distance.

Using the most up-to-date values for $J_{\nu}$ and $D$ that we are aware of, we have with our sign convention of the $J_\nu$'s ($J_\nu>0$ is antiferromagnetic; $J_\nu <0$ is ferromagnetic):  $J_1 = 3.41$ K, $J_2 = -0.14$ K, $J_3 = 0.025$ K and $D = 1.32$ K for \dtox \cite{Yavorskii2008} and $J_1 = 1.56$ K and $D = 1.41$ K for \htox.~\cite{Bramwell2001} Unfortunately, because of the complexity introduced by the large hyperfine coupling interactions in Ho-based materials, much less systematic calorimetric measurements,
 which provide many of the constraints to determine $J_1$ and $J_2$,~\cite{Yavorskii2008}
have been carried out on \hto $ $ compared to \dto $ $. Consequently, the $J_2$ and $J_3$ values for \hto $ $ have not yet been 
determined~\cite{Bramwell2001} and we therefore set $J_2=J_3=0$ for this compound.
 As we shall see below, it turns out that this  ($J_2=J_3=0$)  model describes well the magnetic specific heat of \htox $ $
for the $x=0, 0.4, 0.8 \text{ and } 1.2$ values considered in this work.

For the diluted samples, we assume that the non-magnetic diluting Y$^{3+}$ ions are introduced randomly while {\it all} other parameters of the material, and therefore those of the model in  Eq.~(\ref{gDSM}),  are assumed to be unchanged. This means that, until more accurate microscopic ab-initio modeling of the effect of diamagnetic site-dilution in spin ice compounds becomes available, we ignore local lattice strain effects that may result from the substitution of Dy$^{3+}$ or Ho$^{3+}$ by Y$^{3+}$. In practice we thus ignore any changes that may occur in the $J_\nu$ exchange couplings and the rare-earth ion magnetic moment $\mu$ that would result from variation of the single-ion crystal field ground state wavefunctions.
 This would seem a reasonable first approximation given the close ionic radius of Y$^{3+}$ with Dy$^{3+}$ and Ho$^{3+}$. 
We note in passing that such an approximation has recently been shown to describe quantitatively quite 
well the variation of the critical ferromagnetic temperature in Ho$^{3+}$ substituted by Y$^{3+}$ in LiHo$_{1-x}$Y$_x$F$_4$ all the 
way to,~\cite{Biltmo2007, Biltmo2008} and perhaps even including,
 the dipolar spin glass regime.~\cite{Tam2009,Quilliam2012} In practice, the microscopic $J_\nu$'s and $D$ in Eq.~(\ref{gDSM}) are kept to their pure \dto $ $ and \hto $ $ values while the Ising variables are redefined as $\sigma_i \rightarrow \epsilon_i \sigma_i$, with $\epsilon_i = 0$ if site $i$ is occupied by non-magnetic Y$^{3+}$ ion or $\epsilon_i = 1$ if occupied by a magnetic rare-earth ion. Thus, for [Dy,Ho]$_x$Y$_{2-x}$Ti$_2$O$_7$ the site-random probability distribution of $\epsilon_i$, $P(\epsilon_i)$, is given by $P(\epsilon_i) = (x / 2) \delta(\epsilon_i) + ( 1 - x / 2 ) \delta(\epsilon_i - 1 )$,
where $\delta(u)$ is the Dirac delta function.

\subsection{Monte Carlo Methods}
\label{Sect:MC-methods}

We carried out Monte Carlo simulations for the above model for \dtox $ $ and \htox $ $ at various Y$^{3+}$ 
concentrations $x$. We used a conventional cubic unit cell containing 16 spins, with the system of linear size $L$ having $16 L^3$ spins. Dilution is treated by randomly taking spins out of the system, and a disorder average over 50 
 different random dilution configurations was performed for each dilution level $x$. 
Periodic boundary conditions are used, and we implement the infinite dipole interactions using the Ewald summation technique.~\cite{Wang2001} 
Most of the data production was done with $L=4$ while,
for higher dilutions ($x\geq1.5$),  we used $L=5$ to have a reasonably large number of spins remaining in the system. 
For most of the results presented below, very little system size dependence for the magnetic specific heat, $C_{\rm m}(T)$, data was observed. 

A conventional single spin-flip  
Metropolis algorithm was employed for the Monte Carlo simulation. 
In addition, we used a non-local ``closed-loop''
 update \cite{Melko2001, Melko2004} as well as a new ``open-loop'' update that we now explain. 
The open-loop update is a modified version of the closed-loop update with the following amendments. 
In a diluted system, a fraction of the elementary tetrahedral units will have one or three sites occupied by a spin.
Such ``$\pm$ tetrahedra'' will have the sum of the  Ising $\sigma_i$ variables
 over the occupied sites equal to $\pm 1$ or $\pm 3$. 
At low temperatures, almost all such tetrahedra become constrained to $\pm 1$,  
since these states are energetically lower than the $\pm 3$ ones.

The open-loop update algorithm searches for an end-to-end chain of spins connecting two of these tetrahedra with 
opposite sums of the Ising variables. An open-loop update flips all the spins along the chain when accepted.
Energetically, the nearest-neighbor part of the $\mathcal{H}_{\rm DSIM}$   is 
unchanged in such an open-loop Monte Carlo update. We use the term open-loop update to stress the similarity of the algorithm
 to the original closed-loop update,~\cite{Melko2001, Melko2004} but with the chains of the updated spins ending at 
two ``$\pm 1$ tetrahedra''. In order to further facilitate the equilibrium of the system, we found it necessary to
 also employ the parallel tempering technique which is commonly used in the study of spin glass models.~\cite{Marinari1992}
 At least 200,000 Monte Carlo update steps are used with each single-spin-flip update sweep followed by the two types of loop 
moves update as well as by a parallel tempering replica exchange sweep.~\cite{Marinari1992}
 Another 200,000 such steps are used for data production. 
The magnetic specific heat was determined by performing a disorder average of the energy fluctuations:
\begin{equation}
	C_{\rm m}(T) = { [ { \langle E^2 \rangle - {\langle E \rangle}^2  } ] \over k_{\rm B} T^2 }
\end{equation}
where $\langle ... \rangle$ and $[ ... ]$ are thermal and disorder averages, respectively.

\section{Results and Discussions}

\begin{figure*}[ht]
	\begin{center}
	\includegraphics[width=18cm]{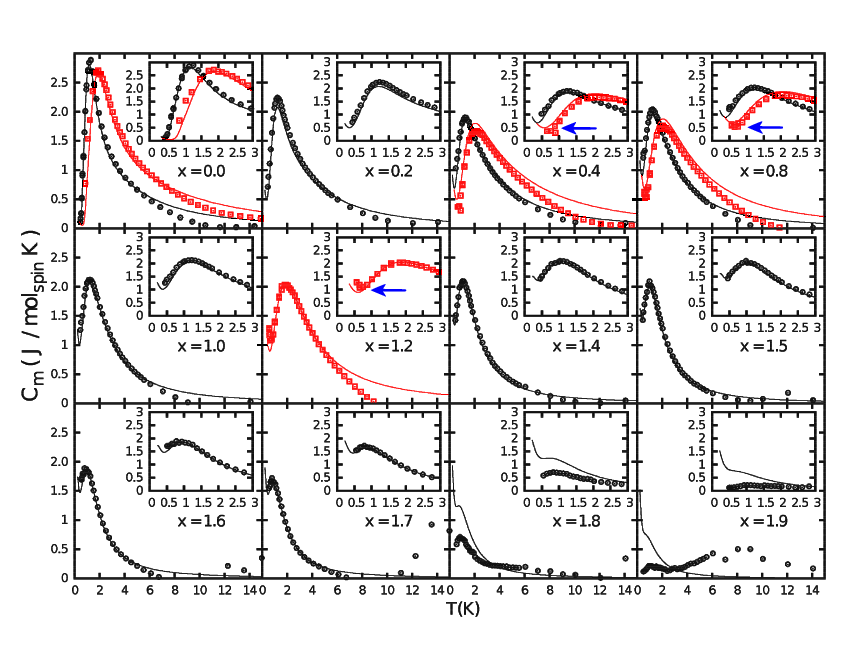}
	\caption{(Color online) Comparison of the magnetic specific heat, $C_{\rm m}(T)$, between Monte Carlo simulations
	and experiments.
	Black open circles are for \dtox $ $ experiment,
	solid black curves are for \dtox $ $ simulations.
	Red open squares are for \htox $ $ experiment,
	and solid red curves are for \htox $ $ simulations.
	Insets show an enlargement around the Schottky peak at $T_{\rm p}$, 
	arising from the formation of the spin ice state.
	The horizontal blue arrows indicate location of $C_{\rm m}(T)$ minima that may be occuring in \htox $ $.
	}
	\label{spe}
	\end{center}
\end{figure*}

We plot in Fig.~\ref{spe} the magnetic specific heat versus temperature, $C_{\rm m}(T)$, obtained 
from Monte Carlo simulations of Eq.~(\ref{gDSM}) (solid lines) for various levels of dilution in comparison 
with experimental data (open black circles for \dtox $ $, open red squares for \htox $ $).

The agreement between our Monte Carlo simulation and the previous experiment \cite{Ke2007} is strikingly good for most dilution levels (up to and including $x=1.7$ for \dtox $ $) and over a rather wide temperature range $T\sim$ [0.4 K $-$ 5 K]. This is particularly noteworthy given that there is {\it no} adjustment of the microscopic parameters of the dipolar spin ice
 Hamiltonian of Eq.~(\ref{gDSM}), except for the dilution of spins in the system. From these results, we can immediately conclude that a simple
site-diluted version of the DSIM of Eq.~(\ref{gDSM})
 does capture the dilution physics of {\it both} materials at a quantitative level. 
This constitutes the main conclusion of this paper.

Close inspection of Fig. \ref{spe} shows that there is a discrepancy in $C_{\rm m}(T)$ between simulation and experimental results for $T\gtrsim 5$ K. Also, the simulation results show a rise of $C_{\rm m}(T)$ as $T$ decreases below a temperature of approximately 0.4 K and 0.6 K for \dtox $ $ and \htox $ $, respectively, while this behavior is barely noticeable in the experimental results. We address these two points in further detail in 
Subsection ~\ref{Sect:high_low}, mostly at the phenomenological level, postponing the discussion of the 
physical implications of these results for the determination of the residual entropy in the following subsection.
In Subsection \ref{Sect:Sres} we present the low temperature limit ($T_0$) dependence of the residual entropy, 
$S_{\rm res}(T_0)$, as a function of dilution level, $x$.
We comment in Subsection \ref{Sect:large-x} 
on the failure of our Monte Carlo simulations  to reproduce the experimental results
for $x=1.8$ and $x=1.9$.

\subsection{High and Low Temperature Regimes}
\label{Sect:high_low}

\subsubsection{High temperature regime}

In the ``high-temperature regime'', typically above 4 K $\sim$ 5 K, we observe that our simulation results for 
 $C_{\rm m}(T)$ depart  from the experimental results. 
Such discrepancies need clarification since 
(i) a demonstration of the validity 
of the microscopic models considered depends on achieving a good degree of 
agreement between experimental and Monte Carlo $C_{\rm m}(T)$ curves and since,
(ii) as we shall see when discussing 
the residual entropy in the next subsection, $C_{\rm m}(T)$ for $T\gtrsim5$ K contributes up to about 10\% 
of the full  ${\text R} \ln(2)$ magnetic entropy.

From a high-temperature expansion perspective, the magnetic specific heat is expected to follow a $C_{\rm m}(T)\sim1/T^2$ form
at temperatures large compared to the typical temperature scale $T_p$,
the temperature at which the specific heat peaks,  set by the interactions in these systems.
This form was indeed verified in all our simulation results.
 In contrast, all the experimental $C_{\rm m}(T)$ data decrease at $T\gtrsim5$ K significantly
faster and are obviously not in agreement with this necessary $1/T^2$ high-temperature form.

We believe this fast drop-off in experiment is likely due to the over-subtraction of the 
lattice contribution to the total specific heat at these temperatures. 
The usual methods for carrying out such a subtraction
rely on an estimated Debye contribution for the acoustic phonons. 
For example, by considering the temperature range of 10 K $\leq T \leq$ 20 K, one might try to fit 
the total specific heat to the form $C_\text{total}(T) = A/T^2 + BT^3$, 
where the $1/T^2$ part comes from the aforementioned magnetic contribution and $T^3$ part is the 
Debye phonon contribution. 
Unfortunately, for $T \gtrsim 10$ K, background contributions from other components 
of the experimental setup become significant. 
 In particular, we note that in order to facilitate thermal conduction in the measurements, 
 Ag powder was mixed into the spin ice powder. At these higher temperatures, the specific heat 
contribution from the Ag powder component becomes larger
 than the magnetic component that we are trying to isolate. 
Fitting the phonon contribution with all these high temperature background contributions 
embeds errors in the $A$ and $B$ fitting parameters, which then causes an over-subtraction 
for the magnetic specific heat $C_m(T)$ at 5 K $\leq T \leq$ 10 K.

\subsubsection{Low temperature regime}

We now turn to the low temperature regime of the $C_{\rm m}(T)$ curves, below the prominent peak at $T=T_p$, with
$T_p \sim 1$ K for \dtox $ $ and $T_p\sim 1.9$ K for \htox $ $. 
In particular,  we discuss the minima found in the simulation results for all dilution levels (including $x=0$, although in this case the minimum 
is more subtle \cite{Melko2001, Melko2004}) in both the Dy and Ho spin ices (see solid curves in insets in Fig. \ref{spe}). 
As discussed in Subsection \ref{Sect:Sres} below, 
the integrated entropy of the system is highly dependent on the $C_{\rm m}(T)$ results at low temperatures since $dS = \frac{C_{\rm m}(T)}{T} dT$.

It is known that in simulations of the undiluted dipolar spin ice model,~\cite{Melko2001, Melko2004} a $C_{\rm m}(T)$
 minimum arises from the development of extra correlations within the spin ice state caused by the dipolar interactions, 
with the system eventually undergoing a transition to long-range order at $T_c \sim 0.13 D$
($T_c \sim 0.18$ K, for the $J_1$, $D$ parameters appropriate for \hto.~\cite{Melko2001, Melko2004})
 For such minima to be found in undiluted spin ice simulations, 
collective spin update algorithms (loop moves discussed in Section \ref{Sect:MC-methods}) have to be included. 
On the other hand, it is very difficult for experiments to display such a 
$C_{\rm m}(T)$ minimum and the long-range order transition, due to the freezing of spins below a temperature $T\sim 0.5$ K.~\cite{Fukazawa2002}

For the diluted systems, the existence of the minima in our simulation suggests that a dynamical arrest similar to the one 
in the undiluted systems does occur.
Indeed, as discussed in Section III B,  equilibrium in  the simulations cannot be 
achieved without using collective update algorithms, further supplemented by
parallel tempering.
 For \dtox $ $, having used a $^{3}$He cryostat (See Section II B),
 the experiments stop at temperatures just above the simulation-predicted minima. 
For \htox $ $, the $C_{\rm m}(T)$ minima 
are perhaps experimentally observed (see horizontal blue arrows in the insets of Fig. \ref{spe}), although the experimental 
data points below the minima do not agree very  well with the simulation results. 
In this case, one should  be warned that there is 
a large nuclear contribution  at $T \lesssim 0.5$ K for Hi$_2$Ti$_2$O$_7$~\cite{Bramwell2001} 
Even though this nuclear component has been subtracted (see Section \ref{Sect:exp-method}), its existence nevertheless
complicates the possible experimental observation of the minima in the magnetic-only part, $C_{\rm m}(T)$, of the total
specific heat $C(T)$.

While the present experimental data do not allow for 
a convincing observation of the minima in $C_{\rm m}(T)$, we unquestionably find them 
in the Monte Carlo simulations of the microscopic DISMs.
The minima observed in the simulations of the diluted DSIMs 
are significantly different from the ones in the undiluted variants.~\cite{Melko2001,Melko2004,Yavorskii2008}
Upon dilution, the $C_{\rm m}(T)$ minimum acquires a significant value, as seen in Fig. \ref{spe}. 
Furthermore, the broad specific heat peak at $T_p(x)$, which signals the development of ice rule correlations
as in the undiluted Dy and Ho spin ices, is less well defined in presence of dilution. 
For example, for $x=1.7$, the peak is more that of a wiggly feature, on the rising $C_{\rm m}(T)$ curve as $T$ approaches zero,
rather than a well-defined peak.   
Indeed, at such a high dilution, the ice rules are marginally 
enforced and the $C_{\rm m}(T)$ peak associated with the development of ice rules fulfilling tetrahedra
is not very prominent.
As discussed further in  Subsection \ref{Sect:Sres} regarding the determination of the residual entropy $S_{\rm res}(T_0)$ 
at a low temperature $T_0$, the behavior of the $C_{\rm m}(T)$ curves suggests that the residual entropy concept employed for  
undiluted spin ices {\it cannot} be readily discussed without a specification of the lowest temperature $T_0$ at which (equilibrated) experimental
data are obtained.

To sum up, there exist significant systematic experimental difficulties in determining the magnetic-only contribution to the specific heat, $C_{\rm m}(T)$, 
in the high temperature regime ($T\gtrsim 5$ K). 
For the low temperature regime ($T\lesssim 0.5$ K), in contrast to the undiluted case, 
the $C_{\rm m}(T)$ curves from our simulations display clear minima with significant $C_{\rm m}(T)$ values. 
On the experimental front, these minima may be marginally observed in \htox $ $ ($x=0.4, 0.8, 1.2$), but are not observed in \dtox $ $. 
At the same time, the very good agreement between the experimental and Monte Carlo $C_{\rm m}(T)$ for both materials (for $x$ up to $x=1.8$ for
\dtox $ $) and for 0.5 K $\lesssim T \lesssim$ 5 K  seemingly vindicates the applicability of a simple site-diluted version of the
DSIM to describe \dtox $ $ and \htox $ $. 
We thus take the following approach.
Having demonstrated good agreement between experiments and models in the temperature range $T\sim$ [0.4 K $-$ 5 K] for both \dtox $ $ and \htox $ $, 
in order to remedy the aforementioned experimental caveats, 
we henceforth only consider the simulation data of Eq.~(\ref{gDSM}) to expose accurately what is the $x$ dependence of the low-temperature
residual entropy, $S_{\rm res}(T_0)$ of the \dtox $ $ and \htox $ $ diluted dipolar spin ice materials.

\subsection{Non-monotonic Residual Entropy}
\label{Sect:Sres}

Since Eq.~(\ref{gDSM}) is an Ising model, the entropy at infinite temperature per mole of spin is ${\rm R} \ln 2$. 
Thus the residual entropy at a given temperature $T_0$ can be written as
\begin{equation}
	S_{\rm res}(T_0) = {\rm R} \ln2 -  \int_{T_0}^\infty { C_{\rm m}(T) \over T } dT
	\label{residual_entropy}
\end{equation}
We plot $S_{\rm res}(T_0)$ obtained from the Monte Carlo simulations for different choices of $T_0$, 
where the integration to infinite temperature
 are done by fitting the $C_{\rm m}(T)$ curves at high temperatures ($>10$ K) to the $1/T^2$ form.

\begin{figure}[h] 
   \centering
   \includegraphics[width=8cm,angle=0]{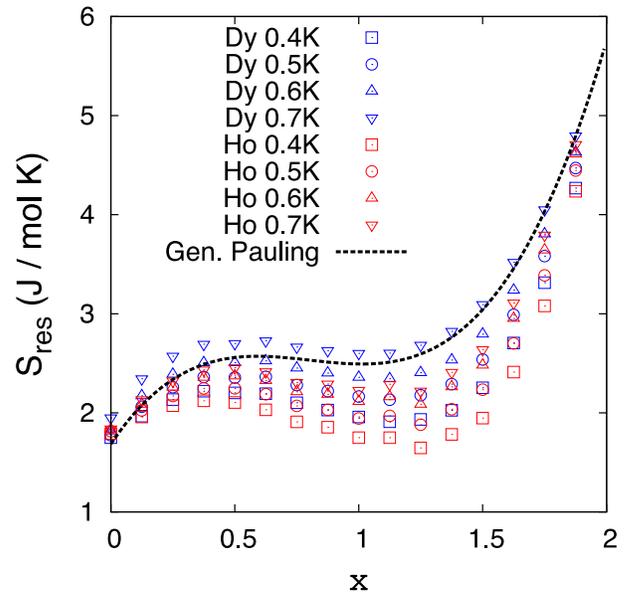}
   \caption{(Color online) Residual entropy determined from Monte Carlo simulations for 
both \dtox $ $ and \htox $ $ with different low temperature
   limits $T_0$. The dotted black curve shows $S_{\rm res}$ given by the generalized Pauling's argument (gPa).
   }
   \label{entropy}
\end{figure}


The results from these Monte Carlo determinations of the residual entropy, $S_{\rm res}(T_0)$ are shown in Fig.~\ref{entropy} for both
\dtox $ $ and \htox $ $. We confirm the previous observation made by Ke {\it et al.} in Ref.~[\onlinecite{Ke2007}]
 that there does exist 
(i) a systematic non-monotonic $x$ dependence of $S_{\rm res}(T_0)$ and (ii)  that there is a difference in $S_{\rm res}(T_0)$ between the two materials.
The main new result here is that, thanks to the ability of the 
 Monte Carlo simulations to provide accurate $C_{\rm m}(T)$ data for $T\lesssim 0.5$ K  and 
$T\gtrsim 10$ K ranges, 
we can now robustly expose both the $x$ dependence and the  materials dependence of $S_{\rm res}$.
Supplementing the previous report,~\cite{Ke2007} we are now also 
uncovering the importance of specifying the base temperature $T_0$ used
in the determination of $S_{\rm res}(T_0)$. Such a need to specify $T_0$ does not arise in previous work on undiluted 
Dy$_2$Ti$_2$O$_7$ and Ho$_2$Ti$_2$O$_7$ because $C_{\rm m}(T)$ practically drops to zero near $T\sim 0.4$ K and $S_{\rm res}$ remains close to the
Pauling value for $C_{\rm m}(T)/T$ integrated upward anywhere from 0.4 K $\pm$ 0.1 K.
In particular, as a final and crucial observation, 
 we note that for all values of $x$ and for a given $T_0$, $S_{\rm res}(x)$ is 
{\it lower} for \htox $ $ than for \dtox $ $, in contrast to the conclusion that
was reached in Ref.~[\onlinecite{Ke2007}] and reproduced in Fig.~\ref{Ke_fig}.

To reiterate, as can be seen in Fig.~\ref{entropy},
the results of the residual entropy for the diluted ($x>0$) DSIM depend strongly on the choice of $T_0$, 
in contrast to the undiluted case ($x=0$), in which the $S_{\rm res}(T_0)$ for different $T_0$s almost collapse onto the calculation of 
the Pauling's entropy, $({\rm R}/2) \ln(3/2)$. For $x=0$, the collapse of the $S_{\rm res}(T_0)$ for different $T_0$'s is the manifestation of 
the {\it projective equivalence} \cite{Isakov2005}, which states that the quasi-ground state properties of the DSIM can be 
described by an effective nearest-neighbor spin ice model up to corrections falling off as $1/r^5$. 
But for $x>0$, the $T_0$ dependence suggests the failure of the projective equivalence upon dilution.



The overall non-monotonic trend of the entropy from the generalized Pauling's argument being in rough qualitative agreement with the results for
the real materials suggests a remnant of the diluted nearest-neighbor spin ice model physics in the diluted DSIMs.
Yet, the two materials, because of their different magnetic interactions, display distinct $S_{\rm res}(x,T_0)$.
Specifically, the two materials possess different energy scales for their 
dipolar interactions, $D$,  relative to the 
nearest-neighbor energy scale, $J_1$ (see Eq.~(\ref{gDSM})).
Thus, the higher overall temperature scale for the formation of
the spin ice state in \htox $ $ compared to \dtox  $ $ results in a residual entropy $S_{\rm res}(T_0,x)$ for \htox $ $ lower 
than for \dtox $ $ for all $x$ and {\it for a given $T_0$}.
However, a choice of $T_0$ that varies for different values of $x$ for a given compound will 
lead to a less smooth $S_{\rm res}(T_0,x)$ evolution than the one seen in Fig.~\ref{entropy} (see Fig.~\ref{Ke_fig}).

\subsection{Large Level of Dilution}
\label{Sect:large-x}

\begin{figure}[h]
   \centering
   \includegraphics[width=8cm,angle=0]{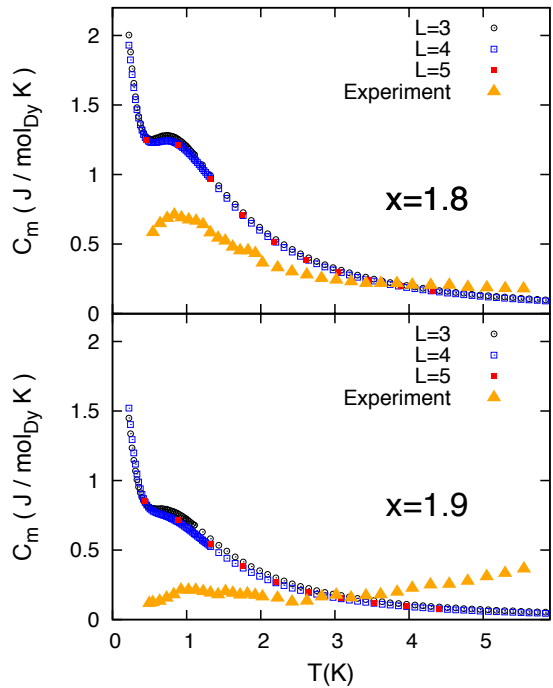}
 \caption{(Color online) 
Comparison of Monte Carlo specific heat with experimental results 
for Dy$_{2-x}$Y$_x$Ti$_2$O$_7$ for sizes $L=3,4,5$ for 
$x=1.8$ (top panel) and $x=1.9$ (bottom panel).
   }
   \label{compare_18_vs_19}
\end{figure}

It is perhaps remarkable that the nice agreement found between Monte Carlo simulations and experiments shown in Fig.~\ref{spe} for \dtox $ $ for
$0 < x \le 1.7$  disappears abruptly and essentially completely 
going from $x=1.7$ to  $x=1.8$ and $x=1.9$ (see Fig.~\ref{spe}).
The only similarity left is that both Monte Carlo and experimental $C_{\rm m}(T)$ data show a small low-temperature hump at 
a temperature $T\sim 0.8$ K that somewhat
agrees between Monte Carlo and experiments (see insets of Fig.~\ref{spe} for $x=1.8$ and $x=1.9$, which 
are further reproduced in Fig.~\ref{compare_18_vs_19}).
This figure further illustrates that despite the large dilution of magnetic ions for $x=1.8$ and $x=1.9$, finite size
effects remain negligible. We are thus rather confident that the discrepancy between simulation 
and experimental results does not arise from computational pitfalls, but is a genuine physical difference.

Presently,  we do not have a good suggestion as to what may cause such a sudden (in terms of ``just'' going from $x=1.7$ to $x=1.8$) 
and large discrepancy between experiments and Monte Carlo data.
A possible mechanism includes the development of a dipolar Ising spin glass state \cite{Tam2009,Quilliam2012} inhibiting thermal equilibrium in the experiments, 
though that should not be at play at temperatures as high as 1 K.
Another possibility includes  a significant random local lattice distortion developing upon reaching 
large levels of dilution. This would affect  the $J_\nu$ couplings and the crystal field, hence the magnetic 
moment $\mu$ and the coupling $D$ compared to the values determined for $x=0$. A third possibility is that of 
a highly uneven distribution of the magnetic ions as $x \rightarrow 2$.
These last two possibilities seem rather unlikely given the close ionic radius of Y$^{3+}$ with Dy$^{3+}$ and Ho$^{3+}$ and the solid solution
that exist in the whole $x \in [0,2]$ range.
More experiments are definitely required to understand the $x \rightarrow 2$ behavior of diluted spin ice materials.

\section{Conclusion}

In this paper we have reported results from Monte Carlo simulations of a site-diluted version of the dipolar spin ice model (DSIM) given by
 Eq.~\ref{gDSM} for \dtox $ $ and \htox. 
A close match between simulation results and experiments in the temperature range 0.5 K $\lesssim T \lesssim$ 5J 
 was found up to, and including, $x=1.7$ (85\% magnetic ions diluted) for \dtox. 
This good agreement between simulations and experiments validates further  the
 underlying dipolar spin ice models for these two compounds.~\cite{Bramwell2001,Yavorskii2008}

The non-monotonicity of the residual entropy as a function of dilution levels, $S_{\rm res}(T_0,x)$,
 is confirmed to originate from the material-specific 
spin-spin interactions themselves, namely the relative strength of the dipolar interactions
with respect to the (mostly) nearest-neighbor exchange coupling $J_1$. 
Furthermore, despite the importance of specifying the base-temperature $T_0$ from which thermodynamic integration of the magnetic specific heat
$C_{\rm m}(T)/T$ is carried out, $S_{\rm res}(T_0,x)$ is nevertheless found to be roughly 
qualitatively described by the generalized Pauling's (gPa) estimate. 
In summary, the difference in the residual entropy $S_{\rm res}$ between 
\dtox $ $ and \htox $ $, as well as with the gPa, have been resolved in the present work.

Encouraged by the robustness of the site-diluted dipolar spin ice 
model to describe the experimental observations for temperatures higher than 0.5 K or so, 
we hope that our work will stimulate further experimental investigations and theoretical studies of spin ice materials at $T \lesssim 0.5$ K, 
in particular in the context of evincing a possible transition to long range order.~\cite{Melko2001,Melko2004}
It would be interesting to explore further the highly diluted regime of \dtox $ $ ($x \ge 1.8$) 
to clarify the origin of the discrepancy between
experimental and Monte Carlo specific heat data  in that regime. It might also be interesting 
to explore the possibility of a dipolar Ising spin glass
state in the highly diluted regime of spin ice materials.~\cite{Tam2009,Quilliam2012}

\begin{acknowledgements}

We thank L. Jaubert, P. McClarty, S. Singer,  P. Stasiak and  K.-M. Tam for helpful discussions. 
This research was funded by the NSERC of Canada and the Canada Research Chair program (M. G., Tier I), 
the Canada Foundation for Innovation (CFI) and the Ontario Innovation Trust (OIT). 
We acknowledge the use of the computing facilities of the Shared Hierarchical 
Academic Research Computing Network (SHARCNET:www.sharcnet.ca). 
P. S. acknowledge support of NSF grants DMR-1104122 and DMR-0701582. 
X. K. gratefully acknowledges the partial financial support by the Clifford G. Shull Fellowship at ORNL.

\end{acknowledgements}


\bibliography{diluted_spin_ice_March23.bib}

\end{document}